# Analysis of proteins in the light of mutations

Jorge A. Vila*

IMASL-CONICET, Universidad Nacional de San Luis, Ejército de Los Andes 950, 5700 San Luis, Argentina.

Proteins have evolved through mutations—amino acid substitutions—since life appeared on Earth, some $10^9$ years ago. The study of these phenomena has been of particular significance because of their impact on protein stability, function, and structure. Three of the most recent findings in these areas deserve to be highlighted. First, an innovative method has made it feasible to massively determine the impact of mutations on protein stability. Second, a theoretical analysis showed *how* mutations impact the evolution of proteins' folding rates. Lastly, it has been shown that native-state structural changes brought on by mutations can be explained in detail by the amide hydrogen exchange protection factors. This study offers a new perspective on how those findings can be used to analyze proteins in the light of mutations. The preliminary results indicate that: (*i*) mutations can be viewed as sensitive probes to identify 'typos' in the amino-acid sequence and also to assess the resistance of naturally occurring proteins to unwanted sequence alterations; (*ii*) the presence of 'typos' in the amino acid sequence, rather than being an evolutionary obstacle, could promote faster evolvability and, in turn, increase the likelihood of higher protein stability; (*iii*) the mutation site is far more important than the substituted amino acid in terms of the protein's marginal stability changes, and (*iv*) the protein evolution's unpredictability at the molecular level— by mutations—exists even in the absence of epistasis effects. Finally, the study's results support the Darwinian concept of evolution as "descent with modification" by demonstrating that some regions of any protein's sequence are susceptible to mutations while others are not. This work contributes to our general understanding of proteins' responses to mutations and may spur significant progress in our efforts to develop methods to accurately forecast changes in protein activity, their propensity for metamorphism, their ability to evolve, and their tendency to aggregate upon mutations.

**Introduction**

State-of-the-art numerical methods have revolutionized structural biology, among other reasons, because they were able to reveal not dozens, hundreds, or thousands, but millions of plausible tridimensional protein structures (Tunyasuvunakool *et al*., 2021; Varadi *et al*., 2022; Callaway, 2022a). This smashing success—despite more than 60 years of continuum research using all-atom force fields (Gibson & Scheraga, 1967; Scheraga, 1968; Lifson & Warshel, 1968; Momany *et al*., 1975; Némethy & Scheraga, 1977; Arnautova *et al*., 2006; Best, 2019) or coarse-grained approaches (Liwo *et al*., 1997; Maisuradze *et al*., 2010; Darré *et al*., 2015; Kmiecik *et al*., 2016; Liwo *et al*., 2021; Jarin *et al*., 2021)—has been followed by new results that include the design of novel proteins (Anishchenko *et al.,* 2021; Callaway, 2022b) among other impressive transformations on structural biology (Wodak, 2023). At this point, it is worth noting that the 'Achilles heel' of these ground-breaking computational approaches—able to solve the protein structure prediction problem (Jumper *et al*., 2021; Cramer, 2021; Serpell *et al*., 2021) but not the protein folding problem (Vila, 2023a; 2023c)—has been, and still is, the need to count with a highly accurate data-set of protein structures solved at an atomic level, where a huge number of parameters can be parametrized (Torrisi *et al*., 2020; Callaway, 2022c); otherwise, these methods will not be able to produce accurate predictions (Vila, 2023a-c). The rationale for the latter is that proteins must be conceived as "analytic wholes" (Vila, 2023a). In line with this requirement, any state-of-the-art numerical method able to accurately predict the effects of mutations on protein stability will require, as a '*sine qua non*' condition, the existence of a large data set where the protein stability changes ($\Delta\Delta G$) have been accurately measured on proteins as a "whole". So far, there is a curated database containing ~8,800 entries that provide accurate information on protein stability changes upon mutations (Xavier *et al*., 2021). Recently, an innovative method has been able to significantly enlarge such a database by making a massive determination of protein stability changes upon mutations in 365 protein domains (Tsuboyama *et al*., 2023). The relevance of this accomplishment is highlighted by the existent demand to solve important problems in structural biology, such as the development of new methods to accurately predict the impact of mutations on protein stability and aggregation, a topic, that has been and still is an object of intensive research (Bloom *et al*., 2006; Zeldovich *et al*., 2007; Capriotti *et al*., 2008; Khan and Vihinen, 2010; Hormoz, 2013; Kulshreshtha *et al*., 2016; Pandurangan and Blundell, 2020; Caldararu *et al*., 2021; Buel and Walters, 2022; Pancotti *et al*., 2022; Pak *et al*., 2023; Diaz *et al*., 2023). The changes in protein stability offer us precise information on the magnitude of the stabilizing or destabilizing effects of mutations. Beyond this obvious conclusion, new questions arise: Is it possible to accurately estimate the impact of mutations on protein evolvability? If so, *how*? For example, is the existence of 'typos' in the sequence of naturally occurring proteins a necessary factor for them to evolve?; *how* resistant are proteins to mutations, and *how* could it affect their evolvability?; or can protein evolution, at the molecular level, be accurately forecast? Before continuing, certain assumptions and definitions need to be made very clear: (*i*) even though there are a variety of biological events that can lead to protein evolution, including natural selection, genetic drift, and migration, we will *only* focus on evolution at the molecular level, namely, by mutations, which are nothing else but the result of nucleotide replacement within a codon (a nucleotide triplet) (Kimura, 1968); (*ii*) we don't distinguish whether a mutation is a result of either protein synthesis (translational error) or gene replication (mutational error) (Epstein, 1966); (*iii*) for any protein amino-acid sequence written in the one-letter code, a 'typo' is defined as one of those letters that, if suitably mutated, will improve the protein's marginal stability and, hence, its chance to evolve (Bloom & Arnold, 2009); (*iv*) the protein 'marginal' stability hereby refers to the Gibbs free-energy gap between the native state and the first unfolded state (Hormoz, 2013; Vila, 2019; Martin & Vila, 2020; Vila, 2021); and (*v*) the analysis will not consider the complex phenomenon of epistasis (Starr & Thornton, 2016; Miton & Tokuriki, 2016; Domingo *et al*., 2019; Miton *et al*., 2020; Miton *et al*., 2021) because it is outside the scope of this work to determine which evolutionary trajectories are most plausible.

Aiming to provide an answer to the above queries, this perspective will start by offering the reader some basic definitions of the factors determining the folding rate evolution upon mutations (Vila, 2023b), as well as *how* the amide Hydrogen eXchange (HX) protection factors can be used to characterize the structural changes that take place as a consequence of mutations (Vila, 2022). Finally, the abundance of proteins whose marginal stability has been precisely determined (Xavier *et al*., 2021; Tsuboyama *et al*., 2023) will enable us to explore the frequency of 'typos' in proteins' amino acid sequences, the resistance of those proteins to unwanted sequence changes, and *how* to characterize the mutation's impact on the protein's ability to evolve in ways that go far beyond those results attained by an analysis in terms of changes in stability alone.

## Results and Discussion

We will explore below *how* and *why* mutations impact different features of proteins and their consequences for their evolution, as well as the possible role of 'typos' in the amino acid sequence as a condition for proteins to evolve.

## Impact of mutation

### *I.- Folding rates and amide HX protection factors*

Mutations—amino acid substitutions—are the mechanisms that nature uses for proteins to evolve. When this happens, the protein stability should vary to some extent and, hence, it will affect the free energy minimum of the protein's native state in a given milieu (the thermodynamic hypothesis). Whether such free-energy change is higher or lower than a threshold value of ~7.4 kcal/mol (Vila, 2021) will determine if the structure remains in its native state or if it will unfold or become nonfunctional (Martin & Vila, 2020; Vila, 2021), *e.g.*, as it does for the single-mutants of the *Aequorea victoria* green fluorescent protein, which loses 100% of the log-fluorescence (native function) if $\Delta\Delta G > \sim 7.5$ kcal/mol (Sarkisyan, *et al*., 2016). The impact of mutations on protein structure and function has been studied for a long time (Zeldovich *et al*., 2007; Stiller *et al.*, 2022), and its analysis will not be revisited here. Rather than this, we will focus on analyzing its impact on the protein's marginal stability (Xavier *et al*., 2021; Vila, 2021; Tsuboyama *et al*., 2023), folding rates (Vila, 2023b), and amide hydrogen exchange protection factors (Vila, 2021; 2022), respectively. In this regard, we have been able to show (Vila, 2023b), using arguments from the theory of state transition, how mutations affect the protein marginal stability ($\Delta\Delta G_m$) and, consequently, the protein folding rates obeying the following relationship:

$$\Delta_\tau = \left(\frac{\tau_m}{\tau_{wt}}\right) \sim e^{\beta\Delta\Delta G_m} \tag{1}$$

where $\beta = 1/RT$, with $R$ the gas constant and T the absolute temperature in Kelvin degrees, and $\tau_x$, with $x = m$ and *wt,* represent the folding rate for the mutant and the wild-type protein respectively, and $\Delta\Delta G_m = (\Delta G_m - \Delta G_{wt})$ the change in the protein marginal stability upon a mutation.

Alternatively, the impact of mutations on protein stability can also be analyzed in terms of the amide Hydrogen eXchange (HX) because it is a sensitive probe with which to assess changes in the protein's native-state structure (Hvidt & Linderstrøm-Lang, 1954; Berger *et al*., 1957; Privalov &Tsalkova, 1979; Huyghues-Despointes *et al*., 1999; Vendruscolo et al., 2003; Craig *et* *al*., 2011; Persson & Halle, 2015; Englander, 2023). This is a conceivable approach because the environment and the native-state structure of the protein both affect intra- and intermolecular hydrogen bonding. Consequently, using arguments from the nuclear magnetic resonance spectroscopy technique, we have been able to show (Vila, 2021; 2022) the following relationship between protein marginal stability changes upon mutations ($\Delta\Delta G_m$) and the amide HX protection factors ($P_f$)—representing the resistance of the amide hydrogen exchange in the native state relative to that of the highest free-energy conformation in the ensemble of folded states—in the EX2 limit (Bahar et al., 1998; Persson & Halle, 2015):

$$\Delta_{P_f} = \left(\frac{P_{f,m}}{P_{f,wt}}\right) \sim e^{\beta\Delta\Delta G_m} \qquad (2)$$

where, $P_x$, with $x = f,m$, and $f,wt$, are the protection factors for the mutant and the wild-type protein, respectively.

Since the changes in folding rates and amide hydrogen-exchange protection factors have the same physical origin—a change in the protein's marginal stability ($\Delta\Delta G_m$)—the similarity between equations (1) and (2) should not surprise us, but it does. Then, a more thorough explanation is needed. Mutations impact Boltzmann's distribution of native-like structures in equilibrium with the lowest Gibbs free energy conformation—the native state—in a way that could affect how that ensemble of structures is capable of preserving its stability (folding rate) and, thereby, its topology (amide hydrogen exchange). At this point, an important question is raised: *why* not accurately forecast the change in the protein's marginal stability upon a mutation? Unfortunately, this option isn't currently possible because the solution to the protein folding problem—that is, *how* the sequence of amino acids encodes its folding—remains unsolved (Moore *et al*., 2022). Moreover, the evidence suggests that an accurate solution to this long-standing problem demands conceiving the proteins as "analytic wholes" and, consequently, solving an *n*-body problem (Vila, 2023a; 2023c). The latter is, after all, the main reason *why* it is so challenging to predict the impacts of mutation (Caldararu *et al*., 2021). Despite this drawback, equations (1) and (2) can be applied to precisely estimate the folding rates and amide HX protection factor changes for proteins based on the determined Gibbs free-energy changes for protein post-mutations. By doing this, we will show the utility of these equations for a better understanding of the underlying structural changes upon mutation. For example, the nature of protein folding rate evolution can be rationalized in terms of protein structural changes that take place after mutations and that could be exquisitely sensed by the amide HX protection factors, which, additionally, may identify barely populated high-energy native-like forms (Vendruscolo *et al*., 2003; Krishna, *et al*., 2004).

### *II.- Assessing the mutation impact beyond protein stability changes*

What are the advantages of using equations (1) and (2) as starting points for our analysis of the impact of mutations rather than just protein stability? Let us explore it. A protein stability change upon a mutation ($\Delta\Delta G_m$) of only ~1.5 kcal/mol—the average strength of protein's hydrogen bonds in solution (Sheu *et al*., 2003)—suggests a 13-fold shift—on either the rate of folding or the amide HX protection factor—when compared to that of the wild-type form. This simple example demonstrates how the protein folding rate and the amide HX protection factors both significantly magnified small free-energy changes, such as those caused by the formation or breaking of intra- or intermolecular hydrogen bonds, that are frequently seen on proteins in solution. Furthermore, their findings may shed important light on the causes of changes in protein structure, evolution, and function. Undoubtedly, by examining the effects of mutations in this way, we can spot information that would otherwise remain hidden if we only looked at minor modifications in protein stability brought on by mutations. To fully comprehend the aforementioned advantages, let's now examine one particular protein—among thousands for which their stability upon mutations has been accurately determined (Tsuboyama *et al*., 2023)—namely the C-terminal domain of NusG protein (PDB ID:2MI6). Each of the 55 residues of this protein was mutated with the remaining 19 naturally occurring amino acids leading to a set of 1,045 mutations (Tsuboyama *et al*., 2023). A detailed graphical analysis in terms of both the stability free-energy change along the structure and their principal component analysis, respectively, was discussed recently (see Figure 3 of Tsuboyama *et al*., 2023). As a complementary study, here, we focus on discussing the information that can be straightforwardly obtained from equations (1) and (2) and how it can be used to easily identify the most sensible residue for destabilizing or stabilizing mutations on the sequence and their impact in terms of the folding rates or amide HX protection factors, respectively. For example, as shown in the inset of Figure **1a** (as filled-magenta squares), mutations of Leu32 will decrease the folding rate ($\tau$) of the mutated protein against that of the wild-type—on average—by ~1,000-fold, while the corresponding maximum stabilizing effect belongs to Pro44 (shown as a blue-filled square)—a residue that lacks an amide group and is located on an antiparallel strand of a *β-sheet*—will increase it by only ~3-fold. It is straightforward to understand the physics underlying this result once it becomes apparent that changes in the amide groups' accessibility to water as a result of mutations at their hydrophobic nucleus (some of them listed in the inset data of Figure **1a** and shown in Figure **1b**) will have a significant impact on the amide HX protection factors. Nothing to be surprised about since it is a well-known phenomenon (Hvidt & Linderstrøm-Lang, 1954; Berger *et al*., 1957; Hvidt & Nielsen, 1966; Privalov &Tsalkova, 1979; Englander *et al*., 1997; Huyghues-Despointes *et al*., 1999; Craig *et al*., 201; Persson & Halle, 2015). The opposite will take place for residues that are already exposed to the solvent, like the polar groups. Overall, the most stabilizing effects caused by mutations in the NusG protein are—on average—three orders of magnitude smaller than the destabilizing ones caused by mutations in the protein's nucleus. Notably, the results from the amide HX protection factors provide evidence that such a fraction of the protein amino-acid sequence is important in modeling both their topology and stability (Dill, 1990).

### III.- Protein evolvability

#### In terms of the protein marginal stability

Let us start with some useful definitions. The change in the protein marginal stability Gibbs free-energy ($\Delta\Delta G_j$) as a consequence of *j*-consecutive mutations ($j > 1$) is a state function (Vila, 2022; 2023b), and hence *all* trajectories in the protein sequence space between a wild-type sequence (*wt*) and that of a target sequence (*ts*) will be indistinguishable (see Figure 3). The latter means that the folding rate of the target sequence ($\tau_{ts}$) will not depend on the trajectory that evolution follows in the sequence space but on the starting folding rate ($\tau_{wt}$) and the final free-energy change $\Delta\Delta G_j$. This is possible because $\Delta\Delta G_j$, as a state function, verifies, for any evolutionary trajectory, the following equality (Vila, 2022):

$$\left(\Delta G_1 - \Delta G_{wt}\right) + \sum_{k=2}^{j}\left(\Delta G_k - \Delta G_{k-1}\right) = \left(\Delta G_j - \Delta G_{wt}\right) = \Delta\Delta G_j \qquad (3)$$

then,

$$\tau_{ts} \sim \tau_{wt}\, e^{\beta\Delta\Delta G_j} \qquad (4)$$

Now, considering that mutations affect the rate at which proteins fold or unfold (Vila, 2023b), an important question arises: Can we distinguish between evolutionary paths such as those shown in Figure 3? From a thermodynamic point of view, the answer is no (see Equation 4), independent of whether intramolecular epistasis—a phenomenon resulting from physicochemical interactions between mutations (Milton *et al*., 2020 and reference therein)—is taken into account or not (Vila, 2022). However, from an evolutionary point of view, the answer is yes, because the time spent to reach the target sequence for any trajectory will depend on both the residue-type mutated and the number of steps (*j*). In other words, *all* evolutionary paths won't be equal. This is a consequence of the fact that each mutation (*m*) could be stabilizing ($\Delta\Delta G_m > 0$), destabilizing ($\Delta\Delta G_m < 0$), or neutral ($\Delta\Delta G_m \sim 0$); then, their impact on the protein folding rate ($\tau_m$) could vary considerably. Therefore, the overall time ($T_j$) needed to go from the wild-type (*wt*) sequence to the target sequence (*ts*) can be defined in terms of each mutational-step (*k*) of a trajectory of *j*-steps, as follows:

$$T_j = \sum_{k=1}^{j} \tau_k \sim \tau_{wt} \sum_{k=1}^{j} e^{\beta \Delta\Delta G_k} \tag{5}$$

where $\tau_k$ is given by Equation (4), namely, $\tau_k \sim \tau_{wt} \, e^{\beta \Delta\Delta G_k}$, with $\Delta\Delta G_k = (\Delta G_k - \Delta G_{wt})$ for each protein evolutionary trajectory in the sequence space (Maynard Smith, 1970). Therefore, the search for the faster—not necessarily the most probable—route demands taking into account the time dependence in reaching the target sequence. Indeed, some of the possible paths might require total times that are biologically forbidden. As we made clear (Vila, 2022), accurately identifying which evolutionary trajectory is the most likely involves taking into account other important contributions, such as epistasis effects (Romero & Arnold, 2009; Sailer & Harms, 2017a). However, this is a topic outside the purview of this work.

A final consideration about the time dependence of the trajectories and the type and number of mutations deserves to be mentioned. Indeed, even if we know exactly the number of steps and the type of mutation for each step of the evolutionary trajectories, the faster pathway will still be unpredictable. The main reason for this is the following: each mutation (*m*) leads to a change in the protein's marginal stability ($\Delta\Delta G_m$), which should be forecast accurately; otherwise, the corresponding folding time for each evolutive step ($\tau_k$) will be subject to uncertainty, as would be the resulting total time ($T_j$). The origin of this problem, as that of intramolecular epistasis, arises from a well-known fact: an answer to the protein folding problem—*how* the amino acid sequence encodes its folding—remains unsolved (Moore *et al*., 2022; Vila, 2023a). The latter also offers an argument to understand the unpredictability of protein evolution as seen from more complex analyses, namely, as a problem that is mainly dependent on unforeseen epistasis effects (Miton & Tokuriki, 2016; Sailer & Harms, 2017b; Domingo *et al*., 2019; Milton *et al*., 2020; Miton *et al*., 2021).

Overall, a straightforward analysis enabled us to demonstrate that the origins of protein evolution's unpredictability—at the molecular level—exist even in the absence of epistasis effects.

### *In the absence and presence of 'typos'*

Let us assume that any naturally occurring proteins possess a 'flawless' amino-acid sequence. This means, from the thermodynamic hypothesis point of view, that the native state of any naturally occurring protein must be in its lowest accessible Gibbs free-energy conformation at a given milieu. Then, any amino acid substitution must necessarily be destabilizing since the stabilizing mutations, that would increase protein marginal stability, are not allowed; otherwise, it would contradict the hypothesis that sequences are 'flawless'. If this were the case, it would have a significant impact on the proteins' ability to evolve because any mutation will lead to a $\Delta G_k \leq \Delta G_{k-1}$ for any $k$, and, hence, $\Delta\Delta G_j$ could quickly become larger than a certain threshold ($\sim -7.4$ kcal/mol) beyond which the protein will unfold and aggregate or become non-functional (Martin & Vila, 2020; Vila, 2021). It should be noted that neutral mutations are also banned since the only allowed neutral mutations would be those with $\Delta\Delta G \equiv 0$, which could only happen if we mutate a residue by itself. Under these assumptions, the protein space (PS) model, where "…*functional proteins must form a continuous network which can be traversed by unit mutational steps without passing through nonfunctional intermediates*…" (Maynard Smith, 1970), shall be a small fraction of the sequence space that proteins should have if 'typos' in the amino-acid sequence were allowed. Indeed, the existence of 'typos' in an amino acid sequence, as opposed to a 'flawless' sequence, would allow the presence of stabilizing mutations as a mechanism for repairing them. Then, the question that arises is twofold: *how* large the protein space could be if mutations can be stabilizing, destabilizing, or neutral, and *what* can we infer from it? Let's make a fast and accurate estimate of the protein space size and its significance for the protein's evolvability. If life began on earth around a billion years ago (Schopf, 2006) and the folding rates for two-state proteins are within the following range: $10^{-3}$ sec to $\sim 1$ sec, independent of protein fold-class, length, or sequence (Vila, 2023b), then the PS would contain between ~$10^{16}$ and ~$10^{19}$ different sequences. The existence of an upper bound cut off on the PS size (~$10^{19}$) means that a 100-residue-length protein would have an average mutation rate per amino acid ($\xi$) of ~1.55 ($\xi^{100}$ ~$10^{19}$) rather than 2 ($\xi^{100}$ ~$10^{30}$), *e.g.*, as a result from modeling the protein as a collection of hydrophobic or polar beads (Lipman & Wilbur, 1991; Bornberg-Bauer, 1997), or 20 ($\xi^{100}$ ~$10^{130}$) as usually assumed (Margoliash & Smith1965; Mandecki, 1998); Dryden *et al*., 2008; Romero & Arnold 2009; Ivankov, 2017); where 20 is the number of naturally occurring amino acids. In other words, the fact that $\xi$ ~1.55 rather than 2 or 20 is equivalent to thinking that a 100-residue-length protein should possess *only* a fraction of its amino acid sequence available to mutations. A conjecture that is supported by the evidence showing that many sites are invariant to mutations in cytochrome *c* (Margoliash & Smith, 1965)—a highly conserved protein consisting of 104 amino acids and presumably utilized ever since the beginning of aerobic life on earth billions of years ago—and by the Darwinian concept of evolution as 'descent with modification' (Darwin, 1859). Indeed, if a portion of a protein's sequence is inaccessible to mutations, it is compatible with that portion of the protein's sequence being 'descent' upon generations, while the remaining portion of the protein's sequence being accessible to mutations is consistent with that portion of the protein's sequence being 'modified' during evolution. Cytochrome *c* of the mammalian type exhibits such a property. Thus, more than 50% of this protein is conserved, demonstrating remarkable evolutionary homology. For example, among the conserved residues are 3 out of 4 prolines, 2 histidines, 2 cysteines, and 11 glycines, with all of those amino acids involved in both the particular topology of the protein as well as in their specific function (Margoliash & Smith, 1965).

The analysis that follows above calls for a response to the following query: *How* often do 'typos' occur in naturally occurring proteins, and *how* do you spot them? The answer to this question will be found in the next section, where the analysis of a large database of mutated naturally occurring proteins will be explored.

### IV.- How resistant to sequence changes are proteins?

According to Anfinsen's dogma (Anfinsen, 1973), the native state, for a given amino-acid sequence, is the thermodynamically most stable structure in a given milieu. Therefore, a mutation represents a sensitive probe to accurately detect 'typos' in the amino acid sequence, *i.e.*, to detect 'sites' where a convenient amino-acid substitution will increase the protein's marginal stability ($\Delta\Delta G_m > 0$). The latter is analogous to claiming that every stabilizing mutation corrects a 'typo' in the amino acid sequence. Moreover, destabilizing mutations provide information about how well the native-state topology is protected against sequence changes. The evidence that is now available suggests that naturally occurring proteins are well protected against mutations, as the majority of them are destabilizing and the percentage of 'typos,' as determined by the number of stabilizing mutations, is far smaller. For example, an analysis of 8,653 proteins based on single mutations (Xavier *et al.*, 2021) shows the following results: ~68% are destabilizing, ~24% are stabilizing, and ~8,0% are neutral mutations (considered here as those with $|\Delta\Delta G|$ changes < 0.1 kcal/mol, since the most conservative mutations, even at the most tolerant sites, frequently lead to $|\Delta\Delta G|$ changes > 0.1 kcal/mol (Matthews, 1995), respectively, while a similar analysis from the observed free-energy distribution from 328,691 out of 341,860 mutations (Tsuboyama *et al.*, 2023), *i.e.*, after removing 13,169 mutations showing $\Delta\Delta G \equiv 0$, which occur when a residue is mutated by itself, indicates that ~71% are destabilizing, ~16% are stabilizing, and ~13% are neutral mutations, respectively. In this regard, Figure 4 shows the dispersion of the destabilizing, stabilizing, and neutral mutations, respectively, as a function of the amino acid type. It should be highlighted that, although their magnitudes are determined by the sites where the mutations occur (as shown in Figure 1), there is no significant variation —on average—in the preference of the amino acids to be neutral, destabilizing, or stabilizing, with the exception—to some extent—of proline (P) and cysteine (C). These exceptions should not surprise us because these two residues possess unique properties among the 20 natural amino acids. For example, cysteine can be present in proteins in reduced or oxidized form (Martin *et al.*, 2010), respectively, making this residue suitable for stabilizing structures; on the contrary, the *cys → trans* isomerization of the peptide groups of the proline residues (Wedemeyer *et al.*, 2002; Vila, *et al.*, 2004) and the lack of an amide-hydrogen, that prevents it from making hydrogen bonding, make this residue a potent breaker of both $\alpha$-helical and $\beta$-sheet structures in globular proteins (Li *et al.*, 1966). Overall, Figure 4's result and Figure 1's illustration of the amide HX protection factor changes due to mutations strongly suggest that, on average, the mutation site, rather than the substituted amino acid, is far more important in terms of the protein marginal stability changes brought by the mutation.

The distribution of the various mutation types found in two distinct data sets, which are compatible with one another, shows that an excess of mutations is destabilizing. The latter highlights the protein's capability to shield its sequence against unwanted mutational effects, such as speeding up its unfolding rate (see Figure 1a), and hence its probability of being degraded—unfolded—at the cellular level. However, not every destabilizing mutation should necessarily be deleterious since the insertion of 'typos' in the protein sequence as a result of this type of mutation could originate—during its evolution—a protein's marginal stability larger than that of the wild-type protein—a feasible possibility, as inferred from Equation 3, and the fact that a less stable protein might provide more rapid turnover—as shown in Equation 4—than more stable proteins.

In addition to all the above, a large number of destabilizing mutations (~70%) would also provide insight into the nature of a protein's native state marginal stability, as revealed by the amide HX protection factor changes. For instance, the perhaps unique arrangement of residues at the protein's, mostly hydrophobic core, may be an indication that any modification to the interactions existent there will ultimately lead to a significant shift in the protein's marginal stability-free energy, which could largely affect its: activity (Giver *et al.*, 1998); propensity for metamorphism (Vila, 2020); folding rate (Vila, 2023b); ability to evolve (Bloom *et al.*, 2006; Bloom & Arnold, 2009); tendency to aggregate (Chiti *et al.*, 2000; Ramirez-Alvarado *et al.*, 2000; Dobson, 2003); *etc*.

## Conclusions

The fact that mRNA provides the amino acid sequence (linked by peptide bonds) that determines the protein's native state enables us to paraphrase Dobzhansky (1973) by saying that "nothing in protein evolution makes sense except in the light of mutations". The latter comes from the fact that a mutation—the origin of evolution at the molecular level—has the possibility of changing the amino acid sequence—which is a hallmark of each protein—and, consequently, the protein's marginal stability, which determines both its folding rate and its capability to evolve. Indeed, an analysis of 340,000 mutations in 365 naturally occurring protein domains enables us to conjecture that the amino acid sequence determines both its capacity to allow 'typos'—as occurs, on average, for ~16% of its sites—as well as its resistance to unwanted mutational effects—as occurs, on average, for ~70% of its sites. Notable, 'typos' may speed up protein evolution rather than prevent it by allowing for a more rapid turnover rate than more stable proteins. This is a feasible phenomenon because mutations affect its folding rates.

The changes in the amide HX protection factors as a result of mutations suggest that the physical forces that protect the native state of proteins determine not only whether a mutation can be stabilizing, destabilizing, or neutral but also whether protein evolvability is predictable or the relevance of where the mutation occurs. Indeed, regardless of the existence of epistasis effects, the entire analysis shows that it is impossible to predict with accuracy *how* a mutation will affect a protein's marginal stability or *how* a protein will evolve through mutations. As explained, the latter is a consequence of the fact that the solution to the protein folding problem—*how* the amino acid sequence encodes its folding—is yet unsolved. In addition, the amide hydrogen exchange protection factors, together with the propensity of the 20 naturally occurring amino acids to be stabilizing, destabilizing, or neutral upon a mutation, indicate that, on average, the precise location where the mutation occurs is more important than the specific amino acid that is substituted. The resulting conclusion is that, while some portions of the protein sequence must be preserved to protect their topology and function, the remaining portions are prone to mutations, allowing the protein to evolve. This is consistent with the Darwinian concept of evolution as "descent with modification."

## Acknowledgment

The author acknowledges support from the IMASL-CONICET-UNSL.

## References

Anishchenko I, Pellock SJ, Chidyausiku TM, Ramelot TA, Ovchinnikov S, Hao J, Bafna K, Norn C, Kang A, Bera AK, DiMaio F, Carter L, Chow CM, Montelione GT, Baker D. De novo protein design by deep network hallucination. Nature. 2021 Dec;600(7889):547-552.

Arnautova YA, Jagielska A., Scheraga H.A. A new force field (ECEPP-05) for peptides, proteins and organic molecules, *J. Phys. Chem. B*. 2006, 110, 5025-5044

Bahar I, Wallqvist A, Covell DG, Jernigan RL.Correlation between native-state hydrogen exchange and cooperative residue fluctuation from a simple model. *Biochemistry* 1998, 37, 1067-1075.

Berger, A.; Linderstrøm-Lang, K. (1957). Deuterium exchange of poly-DL-alanine in aqueous solution. *Arch Biochem Biophys*, 69, 106-118.

Best RB. Atomistic Force Fields for Proteins. *Methods Mol. Biol*. 2019, 2022, 3-19.

Bloom, JD, Arnold, FH. In the light of directed evolution: pathways of adaptive protein evolution. *Proc Natl Acad Sci USA* 2009, 106, 9995-10000.

Bloom, JD, Labthavikul ST, Otey CR, Arnold, FH. Protein stability promotes evolvability. *Proc Natl Acad Sci USA* 2006, 103, 5869-5874.

Bornberg-Bauer E () How are model protein structures distributed in sequence space? *Biophysical Journal* 1997, 73(5), 2393-2403.

Buel GR, Walters KJ. Can AlphaFold2 predict the impact of missense mutations on structure? Nat Struct Mol Biol. 2022 Jan;29(1):1-2.

Caldararu O, Blundell TL, Kepp KP. Three Simple Properties Explain Protein Stability Change upon Mutation. *J Chem Inf Model* 2021, 61,1981-1988.

Callaway E. Alpha Fold's new rival? Meta AI predicts shape of 600 million proteins. *Nature* 2022c, 611, 211-212.

Callaway E. Scientists are using AI to dream up revolutionary new proteins. Nature. 2022b 609, 661-662.

Callaway E. The entire protein universe: AI predicts shape of nearly every known protein. *Nature*. 2022a 608(7921), 15-16.

Capriotti E, Fariselli P, Rossi I, Casadio R. A three-state prediction of single point mutations on protein stability changes. BMC Bioinformatics. 2008 Mar 26;9 Suppl 2(Suppl 2):S6. doi: 10.1186/1471-2105-9-S2-S6. PMID: 18387208.

Chiti F, Taddei N, Bucciantini M, White P, Ramponi G, Dobson CM. Mutational analysis of the propensity for amyloid formation by a globular protein. *EMBO J* 2000, 19, 1441-1449.

Craig, P.O.; Lätzer, J.; Weinkam, P.; Hoffman, R.M.B.; Ferreiro, D.U.; Komives, E.A.; Wolynes, P.G. (2011). Prediction of Native-State Hydrogen Exchange from Perfectly Funneled Energy Landscapes. *J Am Chem Soc*, 133, 17463-17472.

Cramer, P. AlphaFold2 and the future of structural biology. *Nat. Struct. Mol. Biol.* 2021, 28, 704-705.

Darré L, Machado MR, Brandner AF, González HC, Ferreira S, Pantano S. SIRAH: a structurally unbiased coarse-grained force field for proteins with aqueous solvation and long-range electrostatics. *J Chem Theory Comput*. 2015, 11, 723-739.

Darwin, Charles R., 1859, *The Origin of Species by Means of Natural Selection, or the Preservation of Favoured Races in the Struggle for Life*, first edition, London: John Murray.

Diaz DJ, Kulikova AV, Ellington AD, Wilke CO. Using machine learning to predict the effects and consequences of mutations in proteins. *Curr Opin Struct Biol* 2023, 78, 102518.

Dill KA. Dominant forces in protein folding. *Biochemistry* 1990, 29,133-155.

Dobson CM. Protein folding and misfolding. Nature. 2003 Dec 18;426(6968):884-90.

Dobzhansky T. Nothing in Biology Makes Sense except in the Light of Evolution. *The American Biology Teacher* 1973, 35 (3), 125-129.

Domingo J, Baeza-Centurion P, Lehner B. The Causes and Consequences of Genetic Interactions (Epistasis). *Annual Review of Genomics and Human Genetics* 2019, 20, 433-460.

Dryden DTF, Thomson AR, White JH. How much of protein sequence space has been explored by life on Earth? *Journal of the Royal Society Interface* 2008, 5, 953-956.

Englander SW. HX and Me: Understanding Allostery, Folding, and Protein Machines. Annu Rev Biophys. 2023, 52, 1-18.

Epstein CJ. Role of the amino-acid "code" and of selection for conformation in the evolution of proteins. *Nature* 1966, 210, 25-28.

Gibson, K.D.; Scheraga, H.A. (1967). Minimization of polypeptide energy. I. Preliminary structures of bovine pancreatic ribonuclease S-peptide, *Proc Natl Acad Sci USA*, 58, 420-427.

Giver L, Gershenson A, Freskgard PO, Arnold FH. Directed evolution of a thermostable esterase. Proc Natl Acad Sci USA 1998, 95, 12809-12813.

Hormoz, S. Amino acid composition of proteins reduces deleterious impact of mutations. *Sci Rep* 2013, 3, 1-10.

Huyghues-Despointes, B.; Scholtz, J.; Pace, C. (1999). Protein conformational stabilities can be determined from hydrogen exchange rates. *Nat Struct Mol Biol*, 6, 910-912.

Hvidt A, Nielsen SO. Hydrogen exchange in proteins. Adv Protein Chem 1966, 21, 287-386

Hvidt, A.; Linderstrøm-Lang, K. (1954). Exchange of hydrogen atoms in insulin with deuterium atoms in aqueous solutions. *Biochim Biophys Acta*, 14, 574-575.

Ivankov DN. Exact correspondence between walk in nucleotide and protein sequence spaces. *PLoS ONE* 2017, 12(8):e0182525.

Jarin Z, Newhouse J, Voth GA. Coarse-Grained Force Fields from the Perspective of Statistical Mechanics: Better Understanding of the Origins of a MARTINI Hangover. *J Chem Theory Comput*. 2021, 17, 1170-1180.

Jumper *et al*. Highly accurate protein structure prediction with AlphaFold. *Nature* 2021, 596, 583-589.

Khan, S.; Vihinen, M. Performance of protein stability predictors. *Human Mutation* **2010**, *31*, 675-684.

Kimura M (1968) Evolutionary rate at the molecular level. *Nature* 1968, 217, 624-626.

Kmiecik S, Gront D, Kolinski M, Wieteska L, Dawid AE, Kolinski A. Coarse-Grained Protein Models, and Their Applications. *Chem Rev*. 2016, 116, 7898-7936.

Krishna MM, Hoang L, Lin Y, Englander SW. Hydrogen exchange methods to study protein folding. Methods. 2004, 34, 51-64.

Kulshreshtha, S.; Chaudhary, V.; Goswami, G.K.; Mathur, N. Computational approaches for predicting mutant protein stability. *J. Comput. Aided. Mol. Des.* **2016**, *30*, 401-412.

Li SC, Goto NK, Williams KA, Deber CM. Alpha-helical, but not beta-sheet, propensity of proline is determined by peptide environment. *Proc Natl Acad Sci USA* 1996, 93, 6676-6681

Lifson, S.; Warshel, A. (1968). Consistent Force Field for Calculations of Conformations, Vibrational Spectra, and Enthalpies of Cycloalkane and *n*-Alkane Molecules. *The Journal of Chemical Physics*, 49, 5116-5129.

Lipman DJ, Wilbur WJ. Modelling neutral and selective evolution of protein folding. *Proceeding Royal Society London B* 1991, 245, 7-11.

Liwo A, Czaplewski C, Sieradzan AK, Lipska AG, Samsonov SA, Murarka RK. Theory and Practice of Coarse-Grained Molecular Dynamics of Biologically Important Systems. *Biomolecules* 2021, 11, 1347.

Liwo, A., Oldziej, S., Pincus, M.R., Wawak, R.J., Rackovsky, S., Scheraga, H.A. A united-residue force field for off-lattice protein-structure simulations. I. Functional forms and parameters of long-range side-chain interaction potentials from protein crystal data, J. Comput. Chem., 18, 849-873 (1997).

Maisuradze GG, Senet P, Czaplewski C, Liwo A, Scheraga HA. Investigation of protein folding by coarse-grained molecular dynamics with the UNRES force field. *J Phys Chem A*. 2010, 114, 4471-4485.

Mandecki, W.The game of chess and searches in protein sequence space. *Trends Biotechnology* 1998, 16, 200-202.

Margoliash E, Smith EL (1965) Structural and Functional Aspects of Cytochrome *c* in Relation to Evolution, pp. 221-242, in Evolving Genes and Proteins: A symposium. Edited by Vernon Bryson and Henry J. Vogel. Academic Press, New York and London.

Martin OA, Villegas ME, Vila JA, Scheraga HA. Analysis of 13Calpha and 13Cbeta chemical shifts of cysteine and cystine residues in proteins: a quantum chemical approach. J *Biomol NMR* 2010, 46, 217-225.

Martin, AO, Vila, JA.The Marginal Stability of Proteins: How the Jiggling and Wiggling of Atoms is Connected to Neutral Evolution. *Journal Molecular Evolution* 2020, 88, 424-426.

Matthews BW. Studies on protein stability with T4 lysozyme. *Adv Protein Chem*. 1995, 46, 249-78.

Maynard Smith, J. Natural Selection and the concept of a protein space. *Nature* 1970, 225, 563-564.

Miton CM, Buda K, Tokuriki N. Epistasis and intramolecular networks in protein evolution. *Current Opinion in Structural Biology* 2021, 69, 160-168.

Miton CM, Chen JZ, Ost K, Anderson DW, Tokuriki N. Statistical analysis of mutational epistasis to reveal intramolecular interaction networks in proteins. *Methods Enzymol* 2020, 643, 243-280.

Miton CM, Tokuriki N. How mutational epistasis impairs predictability in protein evolution and design. *Protein Sci* 2016, 25, 1260-1272.

Momany FA, McGuire RF, Burgess AW, Scheraga HA. Energy parameters in polypeptides. VII. Geometric parameters, partial atomic charges, nonbonded interactions, hydrogen bond interactions, and intrinsic torsional potentials for the naturally occurring amino acids. *J Phys Chem* 1975, 79, 2361-2381.

Moore PB, Hendrickson WA, Henderson R, Brunger AT. The protein-folding problem: Not yet solved. *Science* 2022, 375, 507.

Némethy G, Scheraga HA. Protein folding. *Q. Rev. Biophys.* 1977, 10, 239-252.

Pak MA, Markhieva KA, Novikova MS, Petrov DS, Vorobyev IS, Maksimova ES, Kondrashov FA, Ivankov DN. Using AlphaFold to predict the impact of single mutations on protein stability and function. *PLoS One* 2023 Mar 16;18(3):e0282689.

Pancotti, C.; Benevenuta, S.; Birolo, G.; Alberini, V.; Repetto, V.; Sanavia, T.; Capriotti, E.; Fariselli, P. Predicting protein stability changes upon single-point mutation: a thorough comparison of the available tools on a new dataset, *Briefings in Bioinformatics* 2022, *23*(2), 1-12.

Pandurangan AP, Blundell TL. Prediction of impacts of mutations on protein structure and interactions: SDM, a statistical approach, and mCSM, using machine learning. *Protein Sci* 2020, 29, 247-257.

Persson, F.; Halle, B. How amide hydrogens exchange in native proteins. *Proc Natl Acad Sci USA* 2015, 112, 10383-10388.

Privalov, P.L.; Tsalkova, T.N. Micro- and macro-stabilities of globular proteins. *Nature* 1979, 280, 693-696.

Ramirez-Alvarado M, Merkel JS, Regan L. A systematic exploration of the influence of the protein stability on amyloid fibril formation in vitro. *Proc Natl Acad Sci USA* 2000, 97, 8979-8984.

Romero PA, Arnold FH, Exploring protein fitness landscapes by directed evolution. *Nat Rev Mol Cell Biol* 2009, 10, 866-876.

Sailer ZR, Harms MJ. High-order epistasis shapes evolutionary trajectories. *PLoS Comput Biol* 2017, 13(5):e1005541

Sailer ZR, Harms MJ. Molecular ensembles make evolution unpredictable. *Proc Natl Acad Sci USA* 2017b, 114, 11938-11943.

Sarkisyan KS, Bolotin DA, Meer MV, Usmanova DR, Mishin AS, Sharonov GV, *et al*. Local fitness landscape of the green fluorescent protein. *Nature* 2016, 533:397-401.

Scheraga, HA. (1968). Calculations of conformations of polypeptides. *Adv Phys Org Chem*, 6, 103-184.

Schopf JW. The first billion years: When did life emerge? Elements 2006, 2, 229-233

Serpell, L.C.; Radford, S.E.; Otzen, D.E. AlphaFold: A Special Issue and A Special Time for Protein Science. *J. Mol. Biol.* 2021, 433, 167231.

Sheu S-Y, Yang D-Y, Selzle HL, Schlag EW. Energetic of hydrogen bonds in peptides. Proc Natl Acad Sci USA 2003, 100,12683–12687.

Starr TN, Thornton JW. Epistasis in protein evolution. *Protein Sci* 2016, 25, 1204-1218.

Stiller JB, Otten R, Häussinger D, Rieder PS, Theobald DL, Kern D. Structure determination of high-energy states in a dynamic protein ensemble. *Nature* 2022, 7901, 528-535.

Torrisi M, Pollastri G, Le Q. Deep learning methods in protein structure prediction. *Comput. Struct. Biotechnol. J.* 2020, 18, 1301-1310.

Tsuboyama K, Dauparas J, Chen J, Laine E, Mohseni Behbahani Y, Weinstein JJ, Mangan NM, Ovchinnikov S, Rocklin GJ. Mega-scale experimental analysis of protein folding stability in biology and design. *Nature* 2023, 620(7973), 434-444.

Tunyasuvunakool, K., Adler, J., Wu, Z. *et al.* Highly accurate protein structure prediction for the human proteome. *Nature* 2021, 596, 590–596.

Varadi M, Anyango S, Deshpande M, Nair S, Natassia C, Yordanova G, Yuan D, Stroe O, Wood G, Laydon A, Žídek A, Green T, Tunyasuvunakool K, Petersen S, Jumper J, Clancy E, Green R, Vora A, Lutfi M, Figurnov M, Cowie A, Hobbs N, Kohli P, Kleywegt G, Birney E, Hassabis D, Velankar S. AlphaFold Protein Structure Database: massively expanding the structural coverage of protein-sequence space with high-accuracy models. Nucleic Acids Res. 2022 50, D439-D444.

Vendruscolo, M; Paci, E.; Dobson, C.M.; Karplus, M. Rare Fluctuations of Native Proteins Sampled by Equilibrium Hydrogen Exchange. *J. Am. Chem. Soc.* **2003**, *125*(51), 15686-15687.

Vila JA, Baldoni HA, Ripoll DR, Ghosh A, Scheraga HA. Polyproline II helix conformation in a proline-rich environment: a theoretical study. *Biophys J* 2004, 86, 731-742.

Vila JA. Thoughts on the Protein's Native State. *J Phys Chem Lett* 2021, 12, 5963-5966.

Vila, JA. Forecasting the upper bound free energy difference between protein native-like structures. *Physica A* 2019, 533, 122053.

Vila, JA. Metamorphic Proteins in Light of Anfinsen's Dogma. *J Phys Chem Lett* 2020, 11, 4998-4999.

Vila, JA. Protein folding rate evolution upon mutations. *Biophys Rev* 2023b, 15, 661-669

Vila, JA. Protein structure prediction from the complementary science perspective. *Biophys Rev* 2023c, 15, 439-445.

Vila, JA. Proteins' Evolution upon Point Mutations. *ACS Omega* 2022, 7, 14371-14376.

Vila, JA. Rethinking the protein folding problem from a new perspective. *Eur Biophys J* 2023a, 52, 189-193.

Wedemeyer WJ, Welker E, Scheraga HA. Proline cis-trans isomerization and protein folding. Biochemistry 2002, 41(50), 14637-14644.

Wodak S. Structural biology: the Transformational Era. *Proteomics* 2023, DOI:10.22541/au.169038361.16607839/v1

Xavier JS, Nguyen TB, Karmarkar M, Portelli S, Rezende PM, Velloso JPL, Ascher DB, Pires DEV. ThermoMutDB: a thermodynamic database for missense mutations. *Nucleic Acids Res*. 2021, 49, D475-D479.

Zeldovich KB, Chen P, Shakhnovich EI. Protein stability imposes limits on organism complexity and speed of molecular evolution. Proc Natl Acad Sci USA 2007, 104,16152-16157.

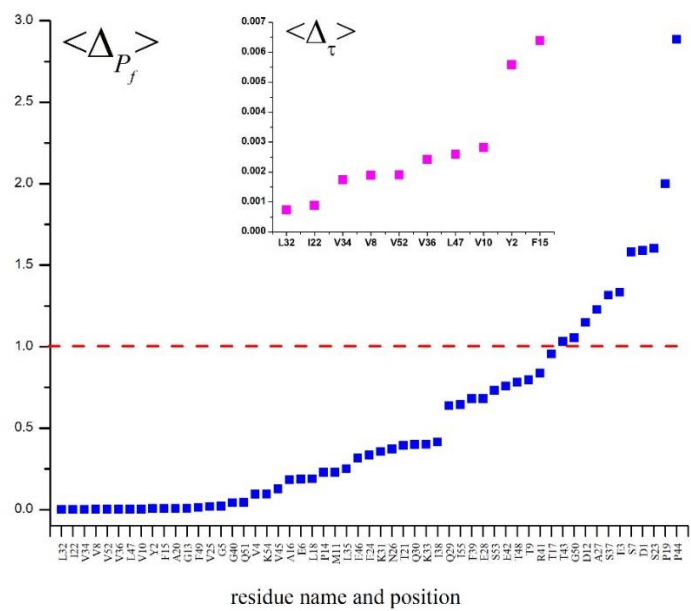

**(a)**

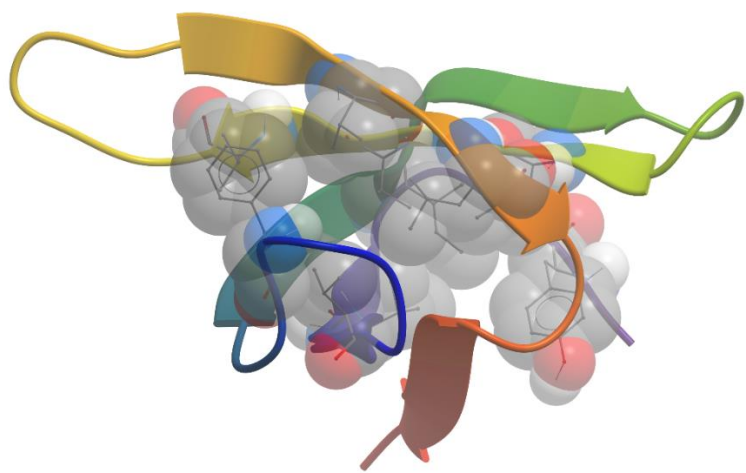

**(b)**

**Figure 1.** (**a**) shows a plot for the median (highlighted as a filled-blue square) of the amide HX protection factor ratio ($< \Delta_{P_f} >$), computed over 19 mutations, as a function of the residue name (in one-letter code) and its position in the sequence of protein 2MI6. The inset shows 10 residues for which the median of the folding rate $<\Delta_\tau>$ (highlighted as filled-magenta squares) leads to the largest protein destabilization, namely for Leu32, which slows down the folding rate of the wild-type by ~1,000-fold. See the text for details. (**b**) show a well-packed core of 10 hydrophobic residues listed in the inset of (**a**).

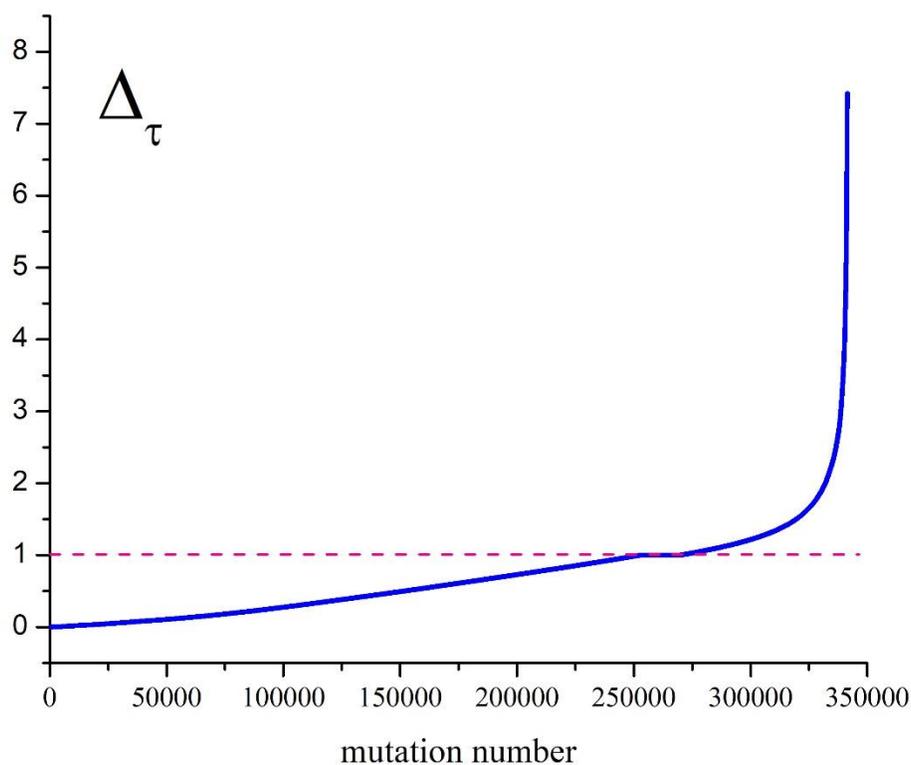

**Figure 2.** The blue line shows the b-spline distribution of the ascending-sorted folding rate ($\Delta_\tau$) computed from equation (1) for 341,400 mutations out of 341,860, *i.e.*, corresponding to 20 mutations on each of the 17,093 residues from 365 protein domains (Tsuboyama *et al.*, 2023). In the graphic, 460 values (~0.1%) possessing a $\Delta\Delta G > 7.4$ kcal/mol were considered outliers (Vila, 2019; 2022) and, hence, are omitted from the figure. The dashed red line highlights the limit between mutations that speed out ($\Delta_\tau < 1$) or speed down ($\Delta_\tau > 1$) the folding rate concerning that of the wild-type. The short plateau existence at $\Delta_\tau \equiv 1$ corresponds to 13,169 mutations showing $\Delta\Delta G \equiv 0$, which occur when a residue is mutated by itself.

(**a**)

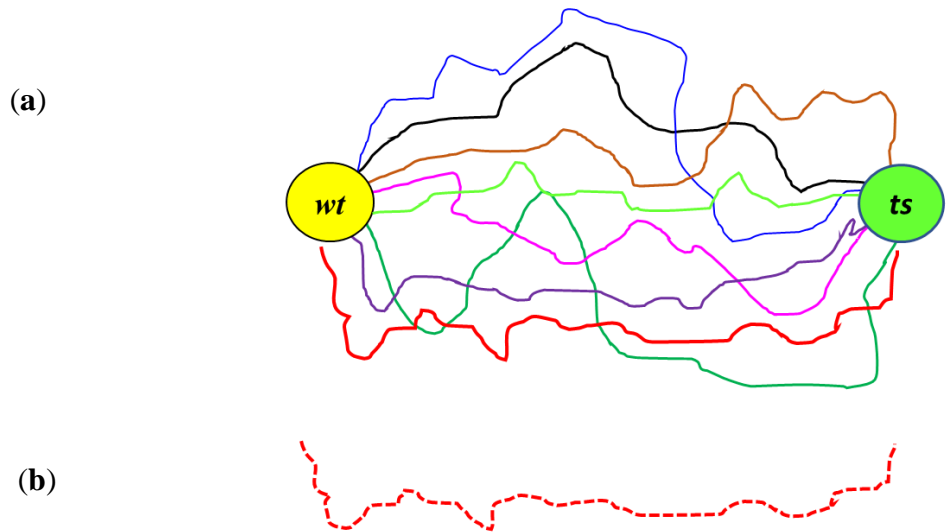

(**b**)

**Figure 3**. (**a**) Cartoon of the protein evolution in the Sequence Space (Maynard Smith, 1970), between a wild-type (*wt*) sequence and the target sequence (*ts*). Lines of various lengths and colors are used to symbolize some of the numerous possible mutational trajectories, each of which has an unidentified number of steps. (**b**) One trajectory, among those shown in (**a**), is isolated to highlight that protein evolution consists of a sequence of steps (shown as dashed-red) where each one designates a functional protein that differs from the previous or next one by one amino acid.

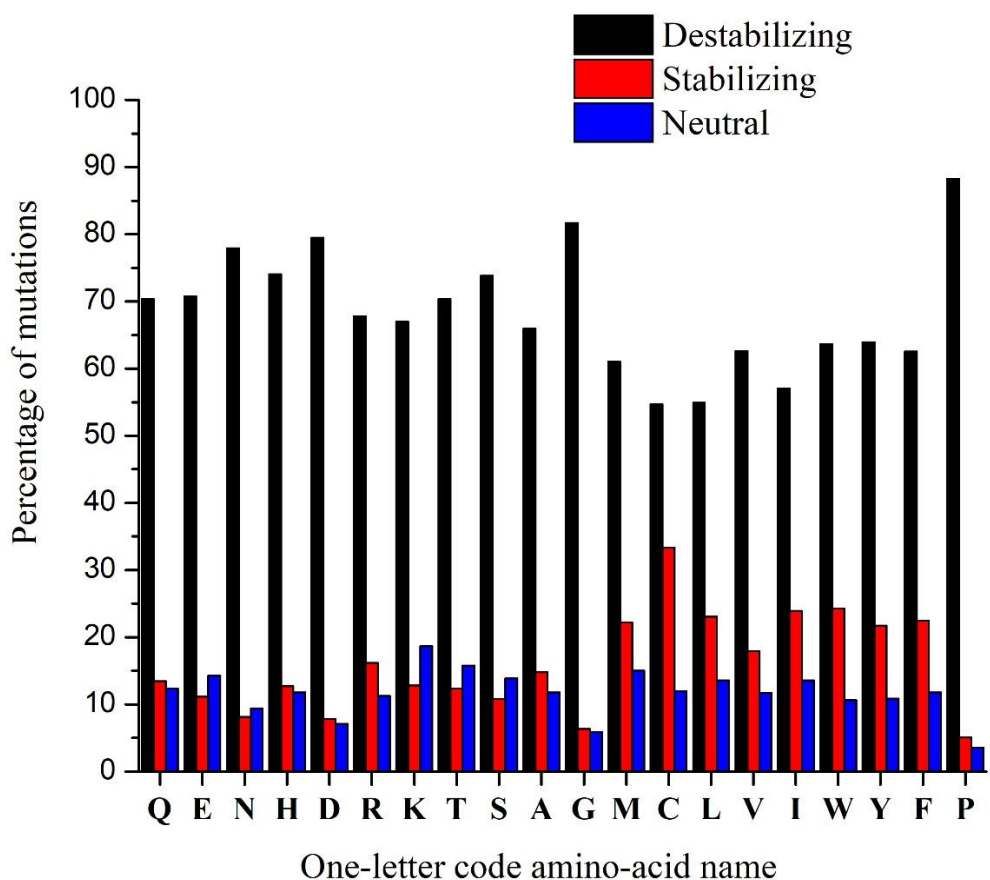

**Figure 4**. Percentage of mutations that are destabilizing (black-filled bars), stabilizing (red-filled bars), and neutral (blue-filled bars) as a function of the amino-acid type. This distribution was obtained from 341,860 mutations, where each of the 20 naturally occurring amino-acids was used on 17,093 sites of a set of 365 protein domains (Tsuboyama *et al.*, 2023).